\documentclass[11pt,a4paper]{article}
\usepackage{jheppub}
\usepackage[utf8]{inputenc}
\usepackage{epstopdf}
\usepackage{tabularx}
\usepackage{amsmath,amssymb, amsfonts,amsbsy,dsfont,mathrsfs,graphicx, booktabs, siunitx, caption, subcaption, slashed}
\usepackage{stmaryrd}
\usepackage{latexsym}
\usepackage{hyperref}
\usepackage{bm}

\newcommand{\eps}{\epsilon}

\newcommand{\nn}{\nonumber}
\newcommand{\lrang}[1]{\langle {#1} \rangle}
\newcommand{\lrbra}[1]{\lbrack {#1} \rbrack}

\newcommand{\lb}{\lbrack}
\newcommand{\rb}{\rbrack}
\newcommand{\la}{\langle}
\newcommand{\ra}{\rangle}

\newcommand{\rar}{\rightarrow}

\newcommand{\beq}{\begin{equation}}
\newcommand{\eeq}{\end{equation}}
\newcommand{\bea}{\begin{eqnarray}}
\newcommand{\eea}{\end{eqnarray}}
\newcommand{\bbm}{\begin{bmatrix}}
\newcommand{\ebm}{\end{bmatrix}}
\newcommand{\bpm}{\begin{pmatrix}}
\newcommand{\epm}{\end{pmatrix}}
\newcommand{\m}{\mu}
\newcommand{\n}{\nu}
\newcommand{\lam}{\lambda}
\newcommand{\lamt}{\tilde{\lambda}}
\newcommand{\etat}{\tilde{\eta}}
\newcommand{\sig}{\sigma}
\newcommand{\sigb}{\bar{\sigma}}
\newcommand{\al}{\alpha}
\newcommand{\alp}{\dot{\alpha}}
\newcommand{\bt}{\beta}
\newcommand{\btp}{\dot{\beta}}
\newcommand{\bb}[1]{\bm{[} #1 \bm{]}}                    % []
\newcommand{\bd}[1]{\bm{\langle}  #1 \bm{\rangle}}      % <>
\newcommand{\bdp}[3]{\left\langle  #1 | #2 |#3 \right] }

\newcommand{\MM}{\mathcal{M}}

\newcommand{\1}{{\bf 1}}
\newcommand{\2}{{\bf 2}}
\newcommand{\3}{{\bf 3}}
\newcommand{\4}{{\bf 4}}
\newcommand{\ib}{{\bf i}}
\newcommand{\jb}{{\bf j}}
\newcommand{\kb}{{\bf k}}
\newcommand{\p}{{\bf p}}

\newcommand{\bchi}{\boldsymbol{\chi}}

\newcommand{\wchi}{\widetilde{\chi}}
\newcommand{\bwchi}{\boldsymbol{\wchi}}

\newcommand{\AddrMainz}{%
PRISMA$^+$ Cluster of Excellence \& Institute of Physics,\\
 Johannes Gutenberg-Universit\"at Mainz, 55099 Mainz, Germany
}

\title{The Rise of SMEFT On-shell Amplitudes}
\author[a]{Rafael Aoude}\emailAdd{aoude@uni-mainz.de}
\affiliation[a]{\AddrMainz}
\author[a]{and Camila S. Machado} \emailAdd{camachad@uni-mainz.de}
\abstract{We present a map between the tree-level Standard Model Effective Theory (SMEFT) in the Warsaw basis and massive on-shell amplitudes. As a first step, we focus on the electroweak sector without fermions. We describe the Feynman rules for a particular choice of input scheme and compare them with the 3-point massive amplitudes in the broken phase. 
 Thereby we fix an on-shell basis which allows us to study scattering amplitudes with recursion relations. We hope  to  open  up new avenues of exploration to a complete formulation of massive EFTs in the on-shell language.}
\preprint{MITP 19-037}

\begin{document}
\maketitle
\newpage
%%%%%%%%%%%%%%%%%%%%%%%%%%%%%%%%%%%%%%%%
\section{Introduction}
The spinor helicity formalism has been an efficient tool to calculate scattering amplitudes with an enormous development over the past decades, especially for the case of massless particles, e.g. \cite{Elvang:2013cua}. The application to massive particles is less explored, although processes involving massive quarks were calculated in~\cite{Dittmaier:1998nn,Forde:2005ue,Rodrigo:2005eu,Schwinn:2006ca,Schwinn:2007ee,Hall:2007mz,Boels:2011zz,Britto:2012qi} and  massive gauge bosons in \cite{Cohen:2010mi,Kiermaier:2011cr,Craig:2011ws}. Recently, a convenient way to describe massive amplitudes, which makes the (massive) little group covariant, was presented in Ref.~\cite{Arkani-Hamed:2017jhn}. This formalism can be applied to supersymmetric theories \cite{Herderschee:2019ofc,Herderschee:2019dmc}, to the Standard Model (SM) \cite{Ochirov:2018uyq,Christensen:2018zcq} and to study amplitudes with higher spin particles \cite{Chung:2018kqs,Guevara:2018wpp,Moynihan:2017tva}. A natural question that follows is: what can we learn in the context of Effective Field Theories (EFTs) using the formalism of massive on-shell amplitudes? 

One way to study the effects of beyond SM (BSM) physics is to write all the possible (independent) higher dimensional operators assuming the SM symmetries. In particular, the \emph{SM Effective Field Theory} (SMEFT) is built on the assumption of a linearly realized electroweak symmetry (for a recent review, see  \cite{Brivio:2017vri}).\footnote{If the electroweak symmetry is non-linearly realized we have the so-called \emph{Higgs Effective Field Theory} (HEFT). We are not exploring this case here \cite{rafacamila}, although it is an interesting problem to understand how the difference between SMEFT and HEFT emerges in the on-shell construction.}~Note that the operators may be related by field redefinitions and finding a basis is, in general, not a trivial task. The so-called Warsaw Basis, for example, was obtained using the equations of motion to remove the maximum number of covariant derivatives~\cite{Grzadkowski:2010es}.\footnote{A general approach involving Hilbert series can be used to find the set of independent operators, see~\cite{Henning:2015daa,Henning:2015alf}. Moreover, a new approach on the construction of a operator basis using the Poincar\'{e}  symmetry of spacetime was presented in~\cite{Henning:2019enq}.}~In addition, the  calculation of observables with the inclusion of higher dimensional operators may involve the computation of diagrams with a large number of external particles and derivatives.

Thus it seems compelling to ask if it is possible to formulate the EFTs without dealing with Lagrangians and equations of motion, and use the spinor language to describe the physics only in terms of on-shell quantities instead. Indeed, this has been already explored, for instance, to calculate QCD amplitudes with the operators  $G^3$ and $hG^2$~\cite{Dixon:2004za,Dixon:1993xd}, to study helicity selection rules \cite{Azatov:2016sqh} and the anomalous matrix \cite{Cheung:2015aba} in the SMEFT (with massless particles), and also for more general EFTs~\cite{Cohen:2010mi,Cheung:2015ota,Cheung:2016drk,Cheung:2018oki,Elvang:2018dco}.

On the other hand, there are several obstacles to extend these results for the SMEFT at low energies, as many of them apply only to massless particles and theories with non-trivial soft limits. Besides, the recursion relations do not always work for EFTs, as contact interactions, which are non-factorizable, are needed as an extra input. There are a few ways to overcome this problem for certain classes of theories~\cite{Cohen:2010mi,Cheung:2015ota,Cheung:2015cba,Elvang:2018dco} and recently, a new strategy was presented  in~\cite{Shadmi:2018xan} to construct amplitudes with higher dimensional operators for an EFT that consists of the SM plus a massive scalar that couples to gluons. This points in the direction of constructing generic EFTs without Lagrangians, fully in the on-shell language.

In this paper, we want to explore the formulation of the massive on-shell SMEFT focusing on the electroweak sector without fermions ($N_f=0$). The eleven dimension six operators can be described in terms of six 3-point on-shell amplitudes: $hhh$, $h\gamma\gamma$, $h\gamma Z$, $hW\bar{W}$, $hZ\bar{Z}$, $W\bar{W}Z$ and $W\bar{W}\gamma$. We explore the fact that all SMEFT deformations at dimension six level can be defined in terms of 3-point on-shell amplitudes. Once this is done, the higher-point amplitudes are already fixed. Independent structures can only appear from dimension eight operators or in a framework where the electroweak symmetry is non-linearly realized. Notice that this is only true because we are not considering the limit of massless particles, in which case, the purely bosonic operators of the Warsaw basis should be defined including 4-point contact interactions~\cite{Ma:2019gtx}. 

Despite the fact that we are not `bootstrapping' the SMEFT from the scattering amplitudes perspective, we hope that this intermediate step can be enlightening and one could see properties that are not so clear in the usual formalism. In addition, we explore the calculation of SM and SMEFT scattering amplitudes with massive particles. For the SM, we calculate the massive amplitude $WW\gamma\gamma$  with the Britto–Cachazo–Feng–Witten (BCFW) recursion relation \cite{Britto:2004ap,Britto:2005fq} and we discuss how to obtain the $WWhh$ demanding correct factorization and unitarity in the high energy limit (HE). For the SMEFT, the recursion relations are challenging and we discuss some possible strategies to overcome the problems. 

The paper is organized as follows: in  Section \ref{s:eft} we discuss how to describe higher dimensional operators with the amplitude language, starting from the massless and moving to the massive case. In Section \ref{s:fr}, we discuss our strategy to do the map of the dimension six operators in the Warsaw basis to the on-shell language which requires defining an appropriate input scheme. Next, in Section \ref{s:amps}, we write the 3-point massive amplitudes and then define the \emph{on-shell basis}. We then move to the discussion about scattering amplitudes using the on-shell basis in Section \ref{s:scattering}.

\section{EFTs with amplitudes}
\label{s:eft}
The massless three point amplitudes are completely fixed by the $U(1)$ Little Group (LG),  locality and by the special 3-point kinematics \cite{Cachazo:2004kj}. This means that there are two non-trivial solutions for the on-shell 3-point amplitudes if we allow the 4-momenta to be complex: the holomorphic (H) with all $\lb i\,j\rb=0$ and the anti-holomorphic (AH) with all $\la i\,j\ra=0$. Hence, the general formula for the 3-point massless amplitudes is given by
\begin{align}
\label{eq:3massless}
	\MM_3(1^{h_1}2^{h_2}3^{h_3})=  g 
	\begin{cases} 
	\bd{12}^{-h_1-h_2+h_3}\bd{23}^{h_1-h_2-h_3} \bd{31}^{h_1-h_2-h_3}\,,~~~~~h<0~\,\text{(H)}, \\
	\,\bb{12}^{h_1+h_2-h_3} \,\,\,\bb{23}^{-h_1+h_2+h_3} \,\bb{31}^{h_1-h_2+h_3}\,,\,~~~~ \, h>0~\,\text{(AH)}, 
	\end{cases}
\end{align}
where $h \equiv \sum_i h_i$.\footnote{In the case of $h=0$, besides the trivial case where all helicities are zero ($\phi^3$-theory),  
it is possible to prove that both cases reduces to a constant of mass dimension one, which vanish for all theories with coupling $[g]\leq 0$. Moreover, these theories fails to fullfill four-points consistency conditions~\cite{McGady:2013sga}.} We can then see that there is a relation between the mass dimension of the coupling $g$ and the allowed vertices. More precisely, a $n$-point amplitude has mass dimension $[\MM_n] =  4 - n$, so using Eq.~(\ref{eq:3massless}) we get that:
\begin{align}
\label{eq:h1g}
|h| = 1-[g] \,.
\end{align} 
In order to connect this formula with terms in the Lagrangian, it is useful to define the complex field strength as \cite{Alonso:2014rga}:~\footnote{In terms of spinor indices the gauge field strength is 
$F_{\mu\nu}\sigma^{\mu}_{\alpha \dot{\alpha}}\sigma^{\nu}_{\beta \dot{\beta}} \equiv F_{\alpha\beta} \bar{\epsilon}_{\dot{\alpha} \dot{\beta}}+ \bar{F}_{\dot{\alpha} \dot{\beta}} \epsilon_{\alpha\beta}$, where $F_{\alpha\beta}$, where  $F_{\alpha\beta}$ ($\bar{F}_{\dot{\alpha} \dot{\beta}}$) corresponding to helicity $+1$($-1$). For a review on spinors, see \cite{Dreiner:2008tw}.}
\begin{align}
\label{e:Xmunu}
X_{\mu\nu}^{\pm} = \frac{1}{2} (X_{\mu\nu } \mp i\tilde{X}_{\mu\nu})\,,\quad \tilde{X}_{\mu\nu}^{\pm}=\pm i X_{\mu\nu}^\pm\,,
\end{align}
with $\tilde{X}_{\mu\nu}=\epsilon_{\mu\nu\alpha\beta}X^{\alpha\beta}/2$. The holomorphic operators are constructed with $X^+$ fields and anti-holomorphic ones with $X^-$. For example, the $\mathcal{O}_X= X^3$ and $\mathcal{O}_{\tilde{X}}=\tilde{X}^3$ operators can be written as $\mathcal{O}_{X}^{\pm} \equiv (\mathcal{O}_X \mp i \mathcal{O}_{\tilde{X}})/2$, such that
\begin{align}
\mathcal{L} \supset c_X \mathcal{O}_X  + c_{\tilde{X}} \mathcal{O}_{\tilde{X}} = c_{X}^{+} \mathcal{O}_{X}^+ +c_{X}^{-} \mathcal{O}_{X}^- \,,
\end{align}
where we define the H and AH complex Wilson coefficients as $c_X^{\pm} = c_X \pm  ic_{\tilde{X}}$.  We can then easily make the connection with Eq.~(\ref{eq:h1g}).  For a coupling with mass dimension $[g]=-2$, the only possible vertices have total helicity $\pm 3$, which are generated by  the H/AH operators $\mathcal{O}_{X}^{\pm}$. For $[g]=-1$, the possible 3-point amplitudes corresponds to the dimension five operators $(X^{\pm})^2\phi$, $\psi^2\phi$ and $\bar{\psi}^2\phi$, where $\psi$ and $\bar{\psi}$ are Weyl spinors.

Moving to higher point amplitudes, we cannot use the 3-point  special kinematics as in this case $s_{ij} = (p_i + p_j )^2$ is non-zero. However, we can still extract some information using the LG scaling and dimensional analysis. In general, a $n$-point tree-level massless amplitude can be written as a sum of angles and square brackets with a common denominator~\cite{Cohen:2010mi}:
\begin{align}
\label{eq:Mn}
\mathcal{M}_n(1^{h_1}\cdots n^{h_n}) \sim g\,\frac{\sum \la \cdots \ra^{n_a}  \lb \cdots \rb^{n_{h}}}{\sum \la \cdots \ra^{d_a}  \lb \cdots \rb^{d_{h}}}\,,
\end{align}
which means that the numerator and denominator are a sum of Lorentz invariant objects. The difference of H/AH contraction is $\Delta_i \equiv n_i - d_i$, then dimensional analysis and LG gives 
\begin{align}
\Delta_a + \Delta_h +[g] = 4-n\,,\quad \Delta_a - \Delta_h = -h \,.
\label{e:contactdim}
\end{align}
For $n>3$, all non-factorizable terms are polynomials in the spinor products. For example, in the case of a $n=4$ with $[g]=-2$, Eq.~(\ref{e:contactdim}) is simply $n_a + n_h =2$, as a non-zero $d_i$ would introduce a pole. For bosonic operators, the only possibility is then $\MM_4 \sim  \{\la \cdots \ra^2, \lb \cdots \rb^2, \la \cdots \ra \lb \cdots \rb \}$. This can be done systematically in order to construct a amplitude basis for higher dimensional operators without make use of a Lagrangian~\cite{Ma:2019gtx}.

However, this simplicity holds just up to dimension six operators as beyond it is challenging to identify the independent terms, as there are terms with more derivatives contributing to the same helicity amplitude. Although arduous to do analytically, Ref.~\cite{Cheung:2016drk} suggests an  numerical approach to identify the redundant terms, which can be useful, for instance, to find a dimension eight basis for the SMEFT.  Another strategy is to explore the spacetime symmetries as in \cite{Henning:2019enq}. 

We are going to pursue a possibility less explored so far, that is to study higher dimensional operators with massive amplitudes. We are then able to study the broken phase of SMEFT amplitudes and define a dimension six basis only using 3-point amplitudes.

For 3-point massive amplitudes, the associated massive LG is the $SO(3) \simeq SU(2)/\mathbb{Z}_2$. The massive momenta can be represented by 4 spinors collected into two vectors $\bchi^I$ and $\bwchi^I$, with $I=1,2$ for the $SU(2)$ indices and greek letters for the Lorentz indices:
\beq
\label{eq:massive}
	p_{\al\btp} = \bchi^I_\al \bwchi_{\btp,I} ~~ \text{with} ~~ \p^2 = \det\bchi \det\bwchi = m^2\,,
\eeq 
where $\det\bchi = \frac{1}{2} \bchi^I\bchi_I$ and $\det\bwchi = \frac{1}{2} \bwchi^I\bwchi_I$. The conventions used here are shown in Appendix~\ref{sec:review} and the massive particles quantities are represented with the {\bf bold} notation, introduced in \cite{Arkani-Hamed:2017jhn}. Moreover, we are going to omit the $SU(2)$ indices for the massive particles and use the short-hand notation as e.g. $\lb {\bf i}\,j \rb^2 \equiv \frac{1}{2} \left( \lb {\bf i}^{I_1}\,j \rb \lb {\bf i}^{I_2}\,j \rb + \lb {\bf i}^{I_2}\,j \rb \lb {\bf i}^{I_1}\,j \rb \right) $. In the rest of the paper, we consider the symmetrization of the $SU(2)$ indices always implicit.

The massive complex momenta respecting the on-shell condition should have $8 - 2 = 6$ real degrees of freedom  (d.o.f). 
Since we introduced 4 two-component spinors, i.e. 16 real  d.o.f, where 2 of them are fixed by Eq.~(\ref{eq:massive}), the remaining 8 d.o.f corresponds to
the $GL(2)$ redundancy $\bchi^J \rar S^J_K \bchi_K$ and $\bwchi_J \rar (S^{-1})^K_J \bwchi_K$,  where $S \in GL(2)$. Whenever the 4-momenta is real, the $GL(2)$ reduces to $SU(2)$ $\otimes$ $U(1)$, 
where the $SU(2)$ part act as rotations on the index $J$ and the $U(1)$ is just a rephasing, as it commutes with the LG. A possible choice used to fix this transformation is given by $\det\bchi = \det\bwchi = m$~\cite{Arkani-Hamed:2017jhn}.\footnote{Since its an arbitrary phase, it is possible to choose to transform, for instance, just the spinor with index $I=2$ as in Ref.~\cite{Conde:2016vxs}. Requiring that the amplitude is invariant under the rephasing may lead to additional constraints on the massive amplitudes. We leave the exploration of this point for a future work.}

The $n$-point amplitudes can then be written as Eq.~(\ref{eq:Mn}), with the difference that now there is also  the $SU(2)$ LG to be satisfied. In particular, the 3-point amplitudes can mix H and AH Lorentz invariants, as $(p_i+p_j)^2$ is non-zero:\footnote{There is a subtley in the case of two particles with same mass, as there is a term in the denominator to fix the LG of the massless particle (for further details, see Appendix~\ref{sec:review}).}
\begin{align}
\label{eq:M3_mass}
\mathcal{M}_3 \,\sim \sum_G g_i \, \la \cdots \ra^{N_a} \lb \cdots \rb^{N_h} \,,
\end{align}
where $G$ is the set of all possible irreducible contractions of the massive/massless spinor variables. (Note that in the massless case $N_a=0$ \emph{or} $N_h=0$ due to the special kinematics.) The size of $G$ is set by the number of the massive external legs and their spinors, we show all possibly cases on the Appendix \ref{sec:review} (where we follow closely Ref.~\cite{Arkani-Hamed:2017jhn} but a different approach to build massive 3-point amplitudes can also be found in \cite{Conde:2016vxs}). Note that dimensional analysis give $N_a+N_h=1-[g_i]$, similarly to the massless case. The difference is that $g_i$ can have powers of masses, so $[g_i]=-2$ does not correspond necessarily to a dimension six operator. 

For a theory with spontaneous symmetry breaking, in the broken phase, we choose to write $g_i$ using powers of $v$ and $m_{W,Z}$ times a dimensionless function, as showed in Table~\ref{eq:F}. As we are going to show in the next section, with these pre-factors, the dimensionless function depend only on the Wilson coefficients $c_i$, the Weinberg angle $\theta$ and the Higgs quartic coupling $ \lambda_h$. We can see that for $[g_i]=\pm 1$ the natural dimensional parameter is the Higgs vev $v$, as the Higgs mechanism shows itself as a IR unification of the UV amplitudes \cite{Arkani-Hamed:2017jhn}. Therefore, 
massless amplitudes apparently unrelated in the UV can be unified into different components of massive amplitudes in the IR. The only case in which no mass or vev should appear is the $[g_i]=-2$ term that corresponds exactly to the $(X^{\pm})^3$ operator, as it is the only possible dimension six operator for a massless 3-point amplitude. For the SM couplings, the HE limit of the amplitudes determines the power of masses (see also \cite{Christensen:2018zcq}). On the other hand, it is unclear how to `bootstrap' the dimensionless functions. For some theories the soft limit of higher point amplitudes can encode information about the symmetries \cite{Elvang:2018dco}, but it is  uncertain how this approach should be applied for the SMEFT. In the next sections, we are going to obtain these functions comparing with the SMEFT Feynman rules.                                      
\renewcommand{\arraystretch}{1.65}
\begin{table}[htb]
\centering
\begin{tabular}{|c||c|c|c|c|}\hline
Coupling dim. &$[g_i]=1$ & $[g_i]=-1$ &  $[g_i]=-2$ \\ \hline\hline
Marginal & $v\, \mathcal{G}_{hhh}$ &  $\frac{1}{v} \,\mathcal{G}_{h\bar{V}V}$ &  $\frac{1}{v\,m_W} \,\mathcal{G}_{W\bar{W}\gamma}\,,$  $\frac{1}{v\,m_W } \,\mathcal{G}_{W\bar{W}Z}\,,\,\frac{1}{v\,m_Z } \,\mathcal{G}_{W\bar{W}Z}$  
 \\ \hline\hline
Dim-$6$ & $\frac{v^3}{\Lambda^2} \,\mathcal{C}_{hhh}$ &  $\frac{v}{\Lambda^2} \,\mathcal{C}_{hV_iV_j}$ &  $\frac{v}{\Lambda^2\,m_W}\, \mathcal{C}_{W\bar{W}V}\,,$ $\frac{1}{\Lambda^2}\, \mathcal{C}_{W\bar{W}V}^{\,\prime}$  
 \\\hline
\end{tabular} 
\caption{The mass dimension of $g_i$ is fixed by the spin/helicities of the 3-point amplitude and we choose to write it as powers of $v$, $m_{W,Z}$ and $\Lambda$ times a dimensionless functions  (we use $\mathcal{G}_i$ for SM  and  $\mathcal{C}_i$ for dimension six structures). In order to do the map to the $N_f=0$ SMEFT we need to write the following 3-point amplitudes: $hhh$, $h\gamma\gamma$, $h\gamma Z$, $hW\bar{W}$, $hZ\bar{Z}$, $W\bar{W}Z$ and $W\bar{W}\gamma$. The kinematic part is fixed by the massless/massive LG. The symmetries and UV properties of the theory are encoded in the dimensionless functions.}\label{eq:F}
\end{table}
\renewcommand{\arraystretch}{1}

\section{The Lagrangian side: Feynman rules and input parameters}
\label{s:fr}
In order to draw the map between the dimension six SMEFT and on-shell amplitudes, we are going to use the Warsaw basis, which is convenient as the number of derivatives are reduced to a minimum in trade for operators with more fields. One advantage is that there are no bivalent operators which makes  the on-shell description more convenient (see also \cite{Cheung:2015aba,Azatov:2016sqh}). The SMEFT Lagrangian is given by
\begin{align}
\mathcal{L}_{\rm SMEFT} = \mathcal{L}_{\rm SM} + \sum_i \frac{c_i}{\Lambda^2} \mathcal{O}_i\,,
\end{align}
where $\Lambda$ is the EFT expansion parameter (from now on we ignore $\Lambda^{-4}$ effects), and the Wilson coefficient $c_i$ and the dimension six operators $\mathcal{O}_i$ can be written in the H and AH form using Eq.~(\ref{e:Xmunu}) (see Table \ref{t:warsaw}). We are considering the operators without fermions and gluons such that the space of higher dimensional operators is 11-dimensional. For the SMEFT Feynman rules we are following the conventions and results of  Ref.~\cite{Dedes:2017zog}. 
\renewcommand{\arraystretch}{1.5}
\begin{table}[htb]
\centering
\begin{tabular}{|c|c||c|c|} \hline
\multicolumn{4}{|c|}{Bosonic operators}  \\ \hline\hline
$\mathcal{O}_{W}^{\pm}$ & $\epsilon^{ijk} W_{\mu\nu}^{i, \pm} W_{\nu\rho}^{j, \pm} W_{\rho\mu}^{k, \pm}  $ &$\mathcal{O}_H$ & $(H^{\dagger}H)^3$ \\
$\mathcal{O}_{HB}^{\pm}$ & $H^{\dagger}H B_{\mu\nu}^{\pm}B_{\mu\nu}^{\pm}$ & $\mathcal{O}_{H\square}$ & $(H^{\dagger}H)\square (H^{\dagger}H)$\\
$\mathcal{O}_{HW}^{\pm}$ & $H^{\dagger}H W_{\mu\nu}^{\pm}W_{\mu\nu}^{\pm}$  &$\mathcal{O}_{HD}$ & $|H^{\dagger}D_{\mu}H|^2$\\
$\mathcal{O}_{HWB}^{\pm}$ & $H^{\dagger}\sigma^i H W_{\mu\nu}^{i,\pm}B_{\mu\nu}^{\pm}$  & &
\\ 
 \hline
\end{tabular} 
\caption{ Bosonic operators in the Warsaw basis. We define the holomorphic and anti-holomorphic operators/Wilson coefficients as $\mathcal{O}_{X}^{\pm} \equiv (\mathcal{O}_X \mp i \mathcal{O}_{\tilde{X}})/2$  and $c_X^{\pm} = c_X \pm  i\tilde{c}_X$. Note that the operators on the right side do not have defined holomorphic structure. }\label{t:warsaw}
\end{table}
\renewcommand{\arraystretch}{1}

Before writing the on-shell amplitudes, one important comment is in order.  The SM has five free parameters which are  $g_L,\, g_Y,\, g_s, \lambda_h$ and $v$, i.e, three gauge couplings, the Higgs quartic coupling and the Higgs vev. A  common choice to extract the values of SM electroweak parameters $g_L$, $g_Y$ and $v$ is through the measured  experimental observables  $\{m_Z, \alpha(0), G_F\}$. These observables will receive contributions from diagrams involving dimension six operators that are usually parametrized as  $\bar{g}_L, \bar{g}_Y, \bar{\lambda}_h, \bar{v} $ where $\bar{x} \equiv \hat{x}+\delta x$ and we are considering just tree-level corrections of order $\mathcal{O}(\Lambda^{-2})$ in the $\delta x$ parameters. To get the bar parameters we have to choose the measurements used to fix the Lagrangian parameters.

From the point of view of the on-shell amplitudes formalism, particle masses constitute
natural input, as they label representations of the Poincar\'{e} group and do not receive any corrections. Since we need to fix four parameters (considering that $g_s$ is fixed via the SM-like triple
gluon self-coupling) and there are only three masses available, we are going to leave the fourth choice unspecified, which means that the formulas will depend explicitly on $\delta v$. Thus, we can define the \emph{on-shell input scheme} as $\{\hat{m}_W, \hat{m}_Z, \hat{m}_h\}$, which follows the SM relations with the hat parameters:
\begin{align}
\hat{m}_W = \frac{\bar{g}_L\, \bar{v}}{2}\,,\qquad \hat{m}_Z = \frac{\sqrt{\bar{g}_L^2+\bar{g}_Y^2} \,\bar{v}}{2}\,,\qquad \hat{m}_h = \bar{\lambda}_h \,\bar{v}^2\,.
\end{align}
Solving these equations we obtain
\begin{align}
 \label{e:input}
& \frac{\delta g_Y}{\hat{g}_Y} = -\frac{\delta v}{\hat{v}}\,-\frac{\hat{v}^2}{\Lambda^2}  \left( c_{HD}\,\frac{(\hat{g}_L^2+\hat{g}_Y^2)}{4\,\hat{g}^2_Y} + c_{HWB}\,\frac{\hat{g}_L}{\hat{g}_Y} \right)\, ,\nonumber \\
 &\frac{\delta \lambda_h}{\hat{\lambda}_h} = - \frac{2\,\delta v}{\hat{v}} +\frac{\hat{v}^2}{\Lambda^2} \left( \frac{3\,c_H}{\hat{\lambda}_h} -2\,c_{H\square} +  \frac{c_{HD}}{2}  \right) \,, \quad \frac{\delta g_L}{\hat{g}_L} = -\frac{\delta v}{\hat{v}}\,.
 \end{align}
One may choose a pseudo-observable for the fourth parameter as the electromagnetic SM-like coupling between the photon and a $W$ pair. Then $\delta v$ can be obtained using the fact  that $\hat{e}\equiv \hat{g}_L\, \hat{g}_Y/\sqrt{\hat{g}_L^2+\hat{g}_Y^2} $ and $\delta g_L$, $\delta g_Y$ of Eq.~(\ref{e:input}), which leads to
 \begin{align}
 \label{e:deltav}
 \delta \tilde{v} \equiv \frac{\delta v \Lambda^2}{v^3} = -\frac{\hat{g}_L^2}{\hat{g}_Y^2}\bigg( c_{HWB}\,\frac{\hat{g}_L \hat{g}_Y}{\hat{g}_L^2+\hat{g}_Y^2}+ \dfrac{c_{HD}}{4} \bigg)\,.
 \end{align}
For now on, we are going to leave $\delta \tilde{v}$ unspecified. The Weinberg angle can then be defined as $c_{\hat{\theta}} \equiv \hat{m}_W/\hat{m}_Z$, which at $\mathcal{O}(\Lambda^{-2})$ gives  $c_{\bar{\theta}} \equiv c_{\hat{\theta}} \,+ \,\delta c_{\theta}$ where,\footnote{One should be careful with this definition as there are extra pieces in the rotation of gauge fields to the mass basis, however, in \cite{Dedes:2017zog}, this is already taken into account in the Feynman rules, and the Weinberg angle can be consistently defined in this way.}
 \begin{align}
 \delta c_{\theta} \equiv \frac{ \hat{g}_Y}{(\hat{g}_L^2+\hat{g}_Y^2)^{3/2}}\,(\hat{g}_Y \delta g_L - \hat{g}_L \delta g_Y)\,.
 \end{align}
Notice that $\delta g_Y, \delta \lambda_h$ and $\delta g_L$ are relevant only in the deviation of SM amplitudes, since whenever a dimension six
Wilson coefficient multiplies them, the term is order $\Lambda^{-4}$. In the following, we drop the hat as it corresponds to the observable quantities.
\section{The on-shell side: SMEFT 3-point amplitudes}
\label{s:amps}
We are going to write the massive 3-point amplitudes for $hhh$, $h\gamma\gamma$, $h\gamma Z$, $hW\bar{W}$, $hZ\bar{Z}$, $W\bar{W}Z$ and $W\bar{W}\gamma$. Comparing with the Feynman rules allows us to draw a 1-to-1 map from on-shell amplitude coefficients and the SM/Warsaw basis parameters.

\subsection{Massive 3-point amplitudes}
In the case of Higgs self-interactions, 
there are no little group indices, hence the most generic on-shell 3-point amplitude is just a constant as we can always trade momentum contractions with masses. The SM and dimension six contributions can be parametrized as
\begin{align}
\mathcal{M}({\bf 1}_{h}{\bf 2}_{h}{\bf 3}_{h}) =  v\, \mathcal{G}_{hhh}\, + \, \frac{v^3}{\Lambda^2}\,\mathcal{C}_{hhh} \,.
\end{align}

Moving to one massive with spin $s$ and two massless particles with helicities $h_2$ and $h_3$, there is a non-trivial constraint on the amplitude given by $|h_3-h_2|\leq s$, which is another way to see the Landau-Yang Theorem \cite{Conde:2016vxs,Arkani-Hamed:2017jhn}. Then, in the case of the coupling of the Higgs with two photons, the only possibility is to have the photons with the same helicity:
\begin{align}
\label{e:hgg}
&\mathcal{M}({\bf 1}_h 2_{\gamma}^{+} 3_{\gamma}^{+})= \frac{v}{\Lambda^2}\,\mathcal{C}_{h\gamma\gamma}^{+} \,\bb{23}^2,
\end{align}
where the amplitude with photons of helicity $-1$ can be obtain just applying a parity transformation, which is equivalent to replacing $\lb2\,3\rb \leftrightarrow \la 2\,3\ra  $ 
and $\mathcal{C}_{h\gamma\gamma}^+ \leftrightarrow \mathcal{C}_{h\gamma\gamma}^-$.

In the case of two massive (with different masses) and one massless particle, the only constraint on the amplitude is $|h_3|\leq s_1 +s_2$.
Then, we can write the amplitude with the Higgs, the $Z$ boson and the photon as
\begin{align}
\label{e:hZg}
&\mathcal{M}({\bf 1}_{h}{\bf 2}_{Z}^{J_{1,2}}3_{\gamma}^{+})=  \frac{v}{\Lambda^2}\,\mathcal{C}_{h Z\gamma}^{+}  \bb{\23}^2\,,
\end{align}
and similarly to the Eq.~(\ref{e:hgg}), the amplitude with the photon with helicity $-1$  can be obtained by parity. 

For the amplitudes with three massive particles the number of terms can be reduced noticing that many structures are zero after the symmetrization of the $SU(2)$ indices.\footnote{This is a manifestation of the Ward identity for massive 
vectors, since $\eps^{K_1K_2}_{\bm 3} \cdot p_{\bm 3} \sim \bdp{\bchi_3}{p_{\bm 3}}{\bwchi_3}^{K_1K_2} = - m_3 \bd{{\bm 3}{\bm 3}}^{K_1K_2} = -m_3\, \varepsilon^{K_1K_3} \rar 0$, where the last step is due the symmetrization of the antisymmetric tensor $\varepsilon^{K_1K_2}$. For massless particles, the Ward identity comes from the fact the brackets with the same particle
vanish, i.e $\bd{33}\rar0$.} The structures with a momentum insertion can be rewritten in terms of contractions of angle and square brackets with the use of Schouten identities (see Appendix \ref{sec:review}). Then, for $V=W,Z$ we have
\begin{align}
\label{e:hVV}
\!\!\!\!\mathcal{M}({\bf 1}_{h}{\bf 2}_V^{I_{1,2}}{\bf 3}_{\bar{V}}^{J_{1,2}})= \!\bigg( \frac{1}{v}\,\mathcal{G}_{hV\bar{V}} + \frac{v}{\Lambda^2}\, \mathcal{C}_{hV\bar{V}} \bigg)\bd{\2\3}\bb{\2\3}
+ \frac{v}{\Lambda^2}\, \mathcal{C}_{hV\bar{V}}^{-} \bd{\2\3}^2  +  \frac{v}{\Lambda^2}\, \mathcal{C}_{hV\bar{V}}^{+}\bb{\2\3}^2 \,,
\end{align}
and the triple gauge coupling $W \bar{W} Z$ can be written as:
\begin{align}
\label{e:WWZ}
\MM (\bm{1}_{W}^{I_{1,2}}\bm{2}_{\bar{W}}^{J_{1,2}}\bm{3}_{Z}^{K_{1,2}}) &= F_1 \,\Big( \bd{\1\2}\bb{\2\3}\bb{\3\1} + \bb{\1\2}\bd{\2\3}\bd{\3\1}\Big) \nn \\
&+ F_2\,\Big( \bd{\3\1}\bb{\1\2}\bb{\2\3} + \bb{\3\1}\bd{\1\2}\bd{\2\3}\Big) \nn \\
&+  F_3\,\Big( \bd{\2\3}\bb{\3\1}\bb{\1\2} + \bb{\2\3}\bd{\3\1}\bd{\1\2}\Big)\, \nonumber \\
&+ \frac{v}{m_W\Lambda^2} \Big(\mathcal{C}_{W\bar{W}Z}^+\, \bd{\1\2}\bb{\2\3}\bb{\3\1} + \mathcal{C}_{W\bar{W}Z}^-\,\bb{\1\2}\bd{\2\3}\bd{\3\1} \Big)\, \nn \\
&+ \frac{1}{\Lambda^2}\Big(\mathcal{C}_{W\bar{W}Z}^{\prime\,+} \bb{\1\2}\bb{\2\3}\bb{\3\1}  +\mathcal{C}^{\prime\,-}_{W\bar{W}Z}  \bd{\1\2}\bd{\2\3}\bd{\3\1} \Big)\,,
\end{align}
where,
\begin{gather}
F_1 = F_2 = \frac{1}{v \,m_Z}  \mathcal{G}_{W\bar{W}Z}+ \frac{v}{m_Z\Lambda^2}\mathcal{C}_{W\bar{W}Z}\,,\quad  F_3 = \frac{1}{v \,m_W}  \mathcal{G}_{W\bar{W}Z}+ \frac{v}{m_W\Lambda^2}\mathcal{C}_{W\bar{W}Z}\,.
\end{gather}
Since all invariants can be written as masses,  e.g. $2 p_\1 \cdot p_\2 = m_3^2 - m_1^2 - m_2^2$, the coefficients of the amplitudes can differ for each term. The form above is particularly convenient to compare with the Feynman rules. Alternatively, one may start with a momentum insertion as $\mathcal{M} \supset (\bd{\1\2}\bb{\1\2}\bdp{\3}{p_{\bm 1}}{\3} + $ cyclic) and use the equations of motion and Schouten identities to write as Eq.~(\ref{e:WWZ}).

Moving to the amplitude of $W \bar{W} \gamma$, we have one massless and two massive particles with same mass and, consequently, only one spinor to construct the amplitude. Following~\cite{Arkani-Hamed:2017jhn}, we can define an auxiliary object $x$ and $x^{-1}$:\footnote{Note that our definition is slightly different from \cite{Arkani-Hamed:2017jhn} since we do not have a mass in the denominator. This is useful to write the amplitude for $W\bar{W}\gamma$ and $W\bar{W}Z$ in a similar form.}
\beq
x \equiv \frac{\bdp{\zeta}{p_2}{3}}{\bd{\zeta3}}\quad\text{and}\quad x^{-1} \equiv \frac{\bdp{3}{p_2}{\zeta}}{\bb{\zeta 3}}\,
\eeq
where $\zeta$ is an arbitrary spinor, $2$ is a massive leg and $3$ is the massless one. They transform as $x \rar t_3^{-2}\, x$ and $x^{-1}\rar t_3^{2}\,x^{-1}$ under the little group of the massless particle and relate to the momenta/polarization vector as:
\begin{align}
x= \sqrt{2} p_1^{\mu} \epsilon_{\mu}^+(p_3) \quad\text{and}\quad x^{-1} = \sqrt{2} p_1^{\mu} \epsilon_{\mu}^-(p_3) \,.
\end{align}
Then, the amplitude can be organized with powers of $x$. This parametrization makes manifest the HE of the interactions, i.e. the minimal coupling appears as the first term in the expansion and it is the one with best UV behaviour (this has interesting applications in the context of black hole physics, see \cite{Chung:2018kqs}). We can also use the relations
\beq
	\bd{\2\1} = \bb{\2\1} + \frac{\bb{\23}\bb{3\1}}{x}\quad \text{and} \quad \bd{\23}\bd{3\1} = - \bb{\23}\bb{3\1}\,\frac{m_W^2}{x^2} \,,
\eeq
to write the amplitude in a similar form as the $W\bar{W}Z$:
\begin{align}
\label{e:WWA}
\mathcal{M}({\bf 1}_{W}^{I_{1,2}}{\bf 2}_{\bar{W}}^{J_{1,2}}3_{\gamma}^{-}) &=  \bigg(\frac{1}{v \,m_W}  \mathcal{G}_{W\bar{W}\gamma}+ \frac{v}{m_W\Lambda^2}\mathcal{C}_{W\bar{W}\gamma}\bigg) \, x^{-1} \bb{\1\2}^2  \nn\\&+ \frac{v\,\mathcal{C}_{W\bar{W}\gamma}^- }{m_W\Lambda^2}\,  \bb{\1\2}\bd{\2 3}\bd{3\1}\,+  \frac{\mathcal{C}_{W\bar{W}\gamma}^{\prime -}}{\Lambda^2}  \bd{\1\2}\bd{\2 3}\bd{3\1} \,. \nn
\end{align}
Having the form of all 3-point amplitudes with gauge bosons allows us to map to the Feynman rules and obtain the dimensionless functions.   We use the convention that  $\mathcal{G}_i$  and $\mathcal{C}_i$ correspond  to the SM-like structures and $\mathcal{G}_i^{\pm}$ correspond only to dimension six operators through H or AH combinations of operators showed in Table~\ref{t:warsaw}. The explicit form of all functions are showed in Tables~\ref{t:sm} and \ref{t:dim6}.

\renewcommand{\arraystretch}{1.35}
\begin{table}[!htb]
\centering
\begin{tabular}{|c||c|c|c|c|}\hline
$\mathcal{G}_i$ & $\mathcal{G}_{hhh}$  & $\mathcal{G}_{hV\bar{V}}$  & $\mathcal{G}_{W\bar{W}Z}$ &$\mathcal{G}_{W\bar{W}\gamma}$  \\ \hline\hline
\text{SM} & $-3\lambda_h$ & $-2$ & $-2\sqrt{2}\, c_{\theta}  $ & $ -2\sqrt{2}\, s_{\theta}$ \\\hline
\end{tabular} 
\caption{Map between the dimensionless $\mathcal{G}_i$ functions and the Standard Model parameters. }\label{t:sm}
\end{table}
\renewcommand{\arraystretch}{1}

\renewcommand{\arraystretch}{1.4}
\begin{table}[!htb]
\centering
\label{tab1}
\begin{tabular}{|c||c|}\hline
$\mathcal{C}_i$ & \text{Warsaw Basis} \\ \hline\hline
$\mathcal{C}_{hhh}$ & $3\lambda_h \delta \tilde{v} + 6 c_H  - 9\lambda_h c_{H\square}+9/4\lambda_h c_{HD}$\\ 
$\mathcal{C}_{W\bar{W}Z}$ & $-4\sqrt{2}\,c_{\theta}\Big[c_{HD}/4 + c_{HWB}\,s_{\theta} c_{\theta}-\delta\tilde{v}\Big] $\\
$\mathcal{C}_{W\bar{W}\gamma}$ & $-\sqrt{2}\, c_{\theta}/t_{\theta}\, \Big[c_{HD}/4 + c_{HWB} s_{\theta} c_{\theta} + \delta \tilde{v}\Big] $ \\
$\mathcal{C}_{hZZ}$ & $-2\Big[ c_{H\square}+c_{HD}/4   -\delta \tilde{v}\Big]$\\ 
$\mathcal{C}_{hW\bar{W}}$ & $-2\Big[c_{H\square} -c_{HD}/4 - \delta\tilde{v}\Big]$\\
\hline \hline
$\mathcal{C}_{h\gamma\gamma}^{\pm}$ & $-2\,\Big[s_{\theta}^2\, c_{HW}^{\pm}  - s_{\theta}\, c_{\theta}\,  c_{HWB}^{\pm} + c_{\theta}^2 \, c_{HB}^{\pm}\Big]$ \\ 
$\mathcal{C}_{hZ\gamma}^{\pm}$ & $-\, \Big[(s_{\theta}^2-c_{\theta}^2\,)\,c_{HWB}^{\pm} + 2 s_{\theta}\, c_{\theta}\, (c_{HW}^{\pm} - c_{HB}^{\pm})\Big]$ \\ 
$\mathcal{C}_{hZZ}^{\pm}$ & $- 2 \Big[ c_{\theta}^2\, c_{HW}^{\pm} + s_{\theta}^2\,c_{HB}^{\pm} + c_{\theta}\,s_{\theta}\,c_{HWB}^{\pm} \Big]$\\
$\mathcal{C}_{hW\bar{W}}^{\pm}$ & $-2\,c_{HW}^{\pm}$\,\\ 
$\mathcal{C}_{W\bar{W}Z}^{\pm}$ &  $ c_{HWB}^{\pm}\sqrt{2}\, s_{\theta} $ \\ 
$\mathcal{C}_{W\bar{W}\gamma}^{\pm}$ & $ c_{HWB}^{\pm} \,\sqrt{2}\, c_{\theta}$\\
$\mathcal{C}^{\prime \pm}_{W\bar{W}Z}$ & $c_W^{\pm}\, 3 \sqrt{2}\,c_{\theta} $  \\
$\mathcal{C}^{\prime \pm}_{W\bar{W}\gamma}$ & $c_{W}^{\pm}\,3\sqrt{2}\,s_{\theta}$ \\\hline
\end{tabular} 
\caption{Map between the dimensionless functions $\mathcal{C}_i,\mathcal{C}_i^{\pm}$ and the dimension six SMEFT in the Warsaw basis, where we define  $\delta \tilde{v}\equiv \Lambda^2\,\delta v/v^3$ (see also Eq.(\ref{e:deltav})).}\label{t:dim6}
\end{table}
\renewcommand{\arraystretch}{1}
\subsection{Defining an on-shell basis}
In addition to the SM couplings, there are 11 parameters in the purely bosonic electroweak sector of SMEFT as can be seen in Table~\ref{t:warsaw}. However, the map from 3-point massive on-shell amplitudes to the Feynman rules has 21 dimensionless functions $\mathcal{C}_i,\mathcal{C}_i^{\pm}$ (see Table~\ref{t:dim6}). This is because the gauge symmetries relate several amplitudes and, in order to define an {\it on-shell basis}, we need to reduce the 21 functions to a 11-dimensional space. There are many equivalent choices but a convenient set that we are going to use to define the \emph{on-shell basis} is given by:
\begin{align}
\mathcal{C}_{hhh}, ~\mathcal{C}_{hZZ},~\mathcal{C}_{hW\bar{W}}, ~\mathcal{C}_{h\gamma\gamma}^{\pm},~\mathcal{C}_{W\bar{W}\gamma}^{\prime \pm},~\mathcal{C}_{W\bar{W}\gamma}^{\pm},~\mathcal{C}_{hW\bar{W}}^{\pm}\,.
\end{align}
The remaining 10 coefficients $(\mathcal{C}_{hZZ}^{\pm},\mathcal{C}_{hZ\gamma}^{\pm}, \mathcal{C}_{W\bar{W}Z}^{\pm}, \mathcal{C}_{W\bar{W}Z}^{\prime \pm}, \mathcal{C}_{W\bar{W}Z}, \mathcal{C}_{W\bar{W}\gamma}$) can be related to the ones of the basis through the following relations:
\begin{gather}
\label{Eq:Frelations1}
	 c^2_\theta\, \mathcal{C}_{hZZ}^{\pm}  = - \sqrt{2} \,s_\theta \,\mathcal{C}_{W\bar{W}\gamma}^{\pm} + s^2_\theta \, \mathcal{C}_{h\gamma\gamma}^{\pm} + (c^2_\theta - s^2_\theta)\,\mathcal{C}_{h\bar{W}W}^{\pm}\,, \nn\\
\,c_\theta\, \mathcal{C}_{W\bar{W}Z}^{\pm} - \, s_\theta\, \mathcal{C}_{W\bar{W}\gamma}^{\pm} = 0\,,  \qquad  \,s_\theta\,\mathcal{C}_{W\bar{W}Z}^{\prime \pm}  - c_\theta\,\mathcal{C}_{W\bar{W}\gamma}^{\prime \pm} = 0\,, \nn\\
2 \,c_\theta s_\theta \, \mathcal{C}_{hZ\gamma}^{\pm}  = 	\mathcal{C}_{h\bar{W}W}^{\pm} -c^2_\theta \, \mathcal{C}_{hZZ}^{\pm} - s^2_\theta\,\mathcal{C}_{h\gamma\gamma}^{\pm}\,,
\end{gather}
and we are still left with two SM-like structures $\mathcal{C}_{W\bar{W}Z}$ and $\mathcal{C}_{W\bar{W}\gamma}$. One linear combination is given by
\beq
	\mathcal{C}_{W\bar{W}Z} + 4\, t_\theta \, \mathcal{C}_{W\bar{W}\gamma} = \Big(\mathcal{C}_{hZZ} -  \mathcal{C}_{hW\bar{W}}\Big)\,2\sqrt{2}\,c_\theta -4\, s_\theta \, c_\theta\, \left(\mathcal{C}_{W\bar{W}\gamma}^{+} + \mathcal{C}_{W\bar{W}\gamma}^{-} \right)\,,
\eeq
and another one is determined once $\delta v$ is fixed.
We can also write the Wilson coefficients $c_i$ of the Warsaw basis in terms of the on-shell basis parameters:
\begin{align}
&c_{W}^{\pm} = \,\frac{\mathcal{C}^{\prime \pm}_{W\bar{W}\gamma}}{3\sqrt{2}\,c_\theta}\,,\qquad  c_{HW}^{\pm} =\,-\frac{\mathcal{C}_{hW\bar{W}}^{\pm}}{2}\,,\qquad c_{HWB}^{\pm} =\,\frac{\mathcal{C}_{W\bar{W}\gamma}^{\pm}}{\sqrt{2}\,c_\theta},\nn \\
& c_{HB}^{\pm}=\, -\frac{1}{2\,c_\theta^2}\left(\mathcal{C}_{h\gamma\gamma}^{\pm} -s_\theta^2 \,\mathcal{C}_{hW\bar{W}}^{\pm} - \sqrt{2}\,s_\theta\, \mathcal{C}_{W\bar{W}\gamma}^{\pm}\right)\,,  \qquad c_H =\,\frac{\mathcal{C}_{hhh}}{6}  - \frac{3\lambda_h \mathcal{C}_{hW\bar{W}}}{4} + \lambda_h\, \delta\tilde{v} , \nn\\
&c_{H\square}= \,-\frac{1}{4}\,\Big(\mathcal{C}_{hW\bar{W}} + \mathcal{C}_{hZZ}\Big) + \delta\tilde{v}\,,\qquad c_{HD}= \Big(\mathcal{C}_{hW\bar{W}} - \mathcal{C}_{hZZ}\Big)  \,.
\end{align}

\section{What about $2\rightarrow 2$ scattering amplitudes?}
\label{s:scattering}

The advantage of recursion relations is well-known in several cases and an important tool for LHC calculations (for a review, see \cite{Elvang:2013cua,Feng:2011np}). In particular, the BCFW recursion relations can be used to calculate massless \cite{Britto:2004ap,Britto:2005fq}  and massive  amplitudes \cite{Badger:2005zh,Badger:2005jv,Schwinn:2007ee} at tree and loop level.
On the other hand, for a general EFT the recursion relations can fail, as normally the amplitudes are not fixed just by factorization.\footnote{One could think that the obstacle for applying recursion relations to EFTs is the energy growth of the amplitudes. However, as discussed in~\cite{Cheung:2015ota}, a bad  high energy behaviour is not an obstacle to the BCFW, as many gravity amplitudes can be calculated correctly in this way. Indeed, in  Ref.~\cite{Cheung:2008dn}, it is shown that the BCFW works for any amplitude in two-derivatives gauge and (super)gravity theories.} Once the information about independent interactions is supplied, the amplitude can be built recursively for a minimum number of external legs. Then, a good large-$z$ behaviour can be obtained with the all-line shift \cite{Cohen:2010mi} or with soft-shifts (for theories with massless particles and non-trivial soft degree)  \cite{Cheung:2015ota,Elvang:2018dco}. 

Another way to calculate amplitudes requiring consistent factorization was presented in \cite{Arkani-Hamed:2017jhn} (see also \cite{Chung:2018kqs}).  This means that for a set of 3-point amplitudes and particle spectrum, one may look for a function that factorizes correctly in each channel, noticing that the residue in one channel can manifest as a pole in another channel. Requiring that the amplitude factorize in all physical channels and that there are no other poles allows to iteratively build the correct amplitude up to polynomial terms (i.e. without poles). 

In order to study massive amplitude with higher dimensional operators we are going to use the BCFW and/or the requirement of consistent factorization channels to build the factorizable part of the amplitude. For the SM, the contact interactions are going to be obtained using information from the high energy limit.  For SMEFT, the 4-point amplitudes are already fixed once the map of the 3-point massive amplitude is done. 
Generically, one may use of the following steps to build massive amplitudes:
\begin{enumerate}
\item Build the $\mathcal{M}_4$ amplitude using recursion relations (e.g. BCFW);
\item Check if the result factorizes correctly in all physical channels. If this is not the case, it means that the recursion relations failed, in the sense that there is a non-zero boundary term;
\item The boundary term can be calculated requiring consistent factorization (see also \cite{Zhou:2014yaa});
\item Once the factorizable part is calculated, write all possible polynomials terms (without poles) consistent with the LG;
\item Compute the HE limit to obtain information about the coefficient of the non-factorizable terms.
\end{enumerate}
Indeed, steps 1-3 can be replaced by simply starting with an ansatz (gluing lower point amplitudes) and building a function that factorizes correctly, as presented in \cite{Arkani-Hamed:2017jhn}. We are going to show a few cases where this algorithm is useful and give the correct result.
\subsection{$W\bar{W}\gamma\gamma$}
Let us start with the SM amplitude $\mathcal{M}^{\rm SM}({\bm 1}_W {\bm 2}_{\bar{W}}3_{\gamma}^{-}4_{\gamma}^{+})$. In this case, the easier choice is to shift the massless legs to the complex plane such that 
\begin{align}
\label{e:shiftspinor}
|\hat{4}\rangle = |4\rangle + z  | 3 \rangle, \qquad |\hat{3} \rbrack = |3\rbrack\, - z | 4 \rbrack\,,
\end{align}
while $|\hat{4}\rbrack = |4\rbrack $ and $|\hat{3}\rangle = |3\rangle$. With the $\lb 3^-, 4^+ \ra$-$\,$shift, there are two factorization channels that corresponds to
\begin{align} 
\hat{p}^2_{\bm q}=(p_{{\bm 1}}+\hat{p}_3(z_t))^2=m_W^2 &\Rightarrow z_t = \frac{2p_{\bm 1}p_3}{\la 3|p_{\bm 1}| 4 \rb }\,, \nonumber \\
\hat{p}^2_{\bm q}=(p_{{\bm 1}}+\hat{p}_4(z_u))^2=m_W^2 &\Rightarrow z_u = -\frac{2p_{\bm 1}p_4}{\la 3|p_{\bm 1}| 4 \rb }\,.
\end{align}
The amplitude can then be represented by
\begin{align}
\mathcal{M}({\bm 1}_W {\bm 2}_{\bar{W}}3_{\gamma}^{-}4_{\gamma}^{+})& = \hat{\mathcal{M}}({\bm 1}_W 3^{-}_{\gamma} {\bm q}_{\bar{W}})\frac{1}{t-m_W^2}\hat{\mathcal{M}}(-{\bm q}_{W} {\bm 2}_{\bar{W}}4^{+}_{\gamma}) \nonumber \\ &+\hat{\mathcal{M}}({\bm 2}_W 3^{-}_{\gamma} {\bm q}_{\bar{W}})\frac{1}{u-m_W^2}\hat{\mathcal{M}}(-{\bm q}_{W} {\bm 1}_{\bar{W}}4^{+}_{\gamma})\nn \\
 &= \hat{\mathcal{M}}_t+\hat{\mathcal{M}}_u\,,
\end{align}
where the $\hat{\mathcal{M}}$ indicate the 3-point amplitudes with the shifted legs of Eq.~(\ref{e:shiftspinor}) corresponding to the $t$ and $u$-channel. We can write the $t$-channel contribution explicitly as 
\begin{align}
\hat{\mathcal{M}}_t = \left( \frac{\mathcal{G}_{W\bar{W}\gamma}}{v\, m_W} \frac{\langle \hat{3}| p_{\bm 1} | \xi_3 \rbrack}{ \lrbra{\xi_3\,\hat{3}}} \lrbra{{\bm 1}\hat{{\bm q}}}^2\right)\frac{1}{t-m_W^2} \left( \frac{\mathcal{G}_{W\bar{W}\gamma}}{ v\, m_W}\frac{\langle \xi_4| p_{\bm 2} | \hat{4} \rbrack}{ \lrang{\hat{4}\,\xi_4}} \lrang{{\bm 2}\,\hat{{\bm q}}}^2\right)\,.
\end{align}
We can choose $\xi_3 = 4$ and $\xi_4 =3$, then the $z$ part is proportional to $\lrang{3\,3} =0$ and $\lrbra{4\,4}=0$ which simplifies the calculation. It is also useful to use momentum conservation to write $\langle 3|p_{\bm 1} | 4 \rbrack =- \langle 3| p_{\bm 2} | 4 \rbrack  $ and 
\begin{align}
\langle{\bm 2}| p_{\bm q} | {\bm 1} \rbrack = \frac{m_W^2}{\langle 3| p_{\bm 2}| 4 \rbrack} \Big( \la 3\, {\bm 1}\ra\lb 4 \, {\bm 2}\rb + \la 3\, {\bm 2}\ra\lb 4 \, {\bm 1}\rb \Big).
\end{align}
The calculation for the $\hat{\mathcal{M}}_u$ is analogous as we can replace $3 \leftrightarrow 4$. We can then sum both contributions to obtain
\begin{align}
\label{e:wwggplus}
\mathcal{M}({\bm 1}_W{\bm 2}_{\bar{W}}3_{\gamma}^{-}4_{\gamma}^{+}) = -\frac{(\mathcal{G}_{W\bar{W}\gamma} \,m_W/v)^{2}}{(t-m_W^2)(u-m_W^2)}\Big( \la 3\, {\bm 1}\ra\lb 4 \, {\bm 2}\rb + \la 3\, {\bm 2}\ra\lb 4 \, {\bm 1}\rb \Big)^2\,,
\end{align}
The square indicates the contraction with the same term with different $SU(2)$ indices and the symmetrization is implicit. Using the map of Table~\ref{t:sm}, we get that $(\mathcal{G}_{W\bar{W}\gamma} \,m_W/v)^{2} = 2e^2$. We can see that naively the amplitude goes as $z^{-2}$, which justifies us  finding the correct result with the chosen shift. However, we should not make use of the final result to prove the large-$z$ behaviour. A more detailed analysis for massive BCFW based on Feynman diagrams (in the case of gluons and massive quarks) can be seen in \cite{Badger:2005zh,Schwinn:2007ee} and it would be interesting to find a generalization for any shift on the lines of \cite{Cheung:2015cba}. 

The good large-$z$ behaviour and the absence of boundary terms can also be seen noticing that an independent contact interaction would have the form
\begin{align}
\label{e:cont}
\frac{\mathcal{G}_{W\bar{W}\gamma\gamma}}{v^4}\Big( \la 3\, {\bm 1}\ra\lb 4 \, {\bm 2}\rb + \la 3\, {\bm 2}\ra\lb 4 \, {\bm 1}\rb \Big)^2\,,
\end{align}
which blows up in  the HE, so indeed $\mathcal{G}_{W\bar{W}\gamma\gamma}=0$. This is the most generic structure consistent with the LG (at  leading order in $1/v$) as any other term with a momentum insertion can be reduced to this one using Schouten identities. 

For the amplitude where the photons have same helicities we can try a different shift where one of the legs is massive.
The $\lb  4^+ ,\bm{1}\ra$-$\,$shift is given by
\begin{align}
\label{e:massshiftspinor}
|\hat{\bm{1}}^I \rangle = |\bm{1}^I\rangle - z \lrang{4\,\bm{1}^I} | 4 \rangle, \qquad |\hat{4} \rbrack = |4\rbrack\, + z \lrang{4\,\bm{1}^K} | \bm{1}_K \rbrack .
\end{align}
Since it is clear how the $SU(2)$ indices should be contracted, we will omit the indices in the following. For this shift, the only factorization channel corresponds to $\hat{p}^2_{\bm q}=(p_{{\bm 1}}+\hat{p}_3(z_t))^2=m_W^2$. (The $u$-channel diagram would have both shifted legs in the same sub-amplitude so it vanishes and there is no $s$-channel as it is not allowed a vertex with photons of the opposite helicities). Solving the condition above for $z_t$, we get
\begin{align}
z_t = \frac{t-m_W^2}{\lrang{3\,4}\langle 4| p_{\bm{1}}| 3 \rbrack} = \frac{\langle 3 |p_{\bm{1}}| 3 \rbrack}{\lrang{3\,4}\langle 4| p_{\bm{1}} | 3 \rbrack}\,.
\end{align}
Then, the amplitude is given by
\begin{align}
\mathcal{M}({\bm 1}_W {\bm 2}_{\bar{W}}3_{\gamma}^{-}4_{\gamma}^{-})& = \hat{\mathcal{M}}({\bm 1}_W 3^{-}_{\gamma} {\bm q}_{\bar{W}})\frac{1}{t-m_W^2}\hat{\mathcal{M}}(-{\bm q}_{W} {\bm 2}_{\bar{W}}4^{-}_{\gamma}) \nn \\
&=\left( \frac{\mathcal{G}_{W\bar{W}\gamma}}{v\,m_W} \frac{\langle 3| \hat{p}_{\bm 1} | \xi_3 \rbrack}{ \lrbra{\xi_3\,3}} \lrbra{{\bm 1}\hat{{\bm q}}}^2\right)\frac{1}{t-m_W^2} \left( \frac{\mathcal{G}_{W\bar{W}\gamma}}{v\,m_W} \frac{\langle 4| p_{\bm 2} | \xi_4 \rbrack}{ \lrbra{\xi_4\,\hat{4}}} \lrbra{{\bm 2}\hat{{\bm q}}}^2\right)
\end{align}
It is convenient to choose $\xi_3 = 4$ and $\xi_4 =3$. We can also use that
\begin{align}
\langle 3| \hat{p}_{\bm 1} | 4 \rbrack &= \langle 3| p_{\bm 1}| 4 \rbrack - z_t \lrang{3\,4} \langle 4| p_{\bm 1} | 4 \rbrack = \frac{m_W^2 s}{\langle 4|p_{\bm 1}| 3 \rbrack} ,
\end{align}
where $\lrang{3\,4}\lrbra{\hat{4}\,3} = m_W^2-u$ and $\lrang{3\,4}\lrbra{3\,4} = s$. This leads to 
\begin{align}
\label{e:wwggminus}
\mathcal{M}({\bm 1}_W {\bm 2}_{\bar{W}}3_{\gamma}^{-}4_{\gamma}^{-}) = \frac{(\mathcal{G}_{W\bar{W}\gamma} \,m_W/v)^2}{(t-m_W^2)(u-m_W^2)}\lb  {\bm 1}\, {\bm 2}\rb ^2 \la 3\,4\ra^2  \,,
\end{align}
which agrees with the Feynman rules calculation. Naively, we can see that with the one massless/one massive shift the amplitude scales as $z^{-2}$. On the other hand, a shift on both massless legs would go  as $z^0$, which explains our choice. In this case, an independent contact interaction is  also forbidden due to the HE constraint. 

The next step is to calculate the amplitude  with one insertion of a dimension six operator $\mathcal{M^{\rm BSM}}({\bm 1}_W {\bm 2}_{\bar{W}}3_{\gamma}^{-}4_{\gamma}^{+})$.  The  SM-like part  gives the same result as Eq.~(\ref{e:wwggplus}), with the appropriate  couplings.  For the non-minimal coupling in the vertex $\hat{\mathcal{M}}(-{\bm q}_{W} {\bm 2}_{\bar{W}}4^{-}_{\gamma})$, we can use the $\lb 3^{-},4^{+}\rangle$-shift of Eq.~(\ref{e:shiftspinor}). One of the 3-point amplitudes corresponds to the non-minimal coupling of Eq.~(\ref{e:WWA}) and we obtain:
\begin{align}
&\hat{\mathcal{M}}^{{\rm BSM},-}_t = \bigg(\frac{1}{v \,m_W}  \mathcal{G}_{W\bar{W}\gamma}+ \frac{v}{m_W\Lambda^2}\mathcal{C}_{W\bar{W}\gamma}\bigg)\frac{m_W^2}{s\,(t-m_W^2)}  \times  \\
&\times \left\{  \frac{v\,\mathcal{C}_{W\bar{W}\gamma}^- }{\Lambda^2}   \Big( \la 3\, {\bm 1}\ra\lb 4 \, {\bm 2}\rb + \la 3\, {\bm 2}\ra\lb 4 \, {\bm 1}\rb \Big)\,\la 3\, {\bm 1}\ra \la 3\, {\bm 2}\ra  \lb 4\,3\rb\,    + \frac{\mathcal{C}^{\prime -}_{W\bar{W}\gamma}}{\Lambda^2}  \frac{\la {\bm 1}\,{\bm 2}\ra \lb 4\,3 \rb}{m_W} \la 3\,{\bm 1} \ra \la 3\,{\bm 2}\ra  \la 3| p_{\bm 2}|4 \rb \right\}\,. \nn
\end{align}
Then, $\hat{\mathcal{M}}^{{\rm BSM},+}_t$ can be obtained by parity and the $u$-channel using crossing symmetry. Summing the $t$ and $u$ contributions and keeping just $\mathcal{O}(\Lambda^{-2})$ terms leads to
\begin{align}
\label{e:comptonbsm1}
&\mathcal{M}^{{\rm BSM},-}({\bm 1}_W {\bm 2}_{\bar{W}}3_{\gamma}^{-}4_{\gamma}^{+}) =  \bigg(\frac{\mathcal{G}_{W\bar{W}\gamma}}{\Lambda^2}   \bigg)\frac{1}{(t-m_W^2)\,(u-m_W^2)} \times \\
&\times \left\{ \mathcal{C}_{W\bar{W}\gamma}^- m_W  \Big( \la 3\, {\bm 1}\ra\lb 4 \, {\bm 2}\rb + \la 3\, {\bm 2}\ra\lb 4 \, {\bm 1}\rb \Big) \la 3\, {\bm 1}\ra \la 3\, {\bm 2}\ra  \lb 4\,3\rb  +\, \mathcal{C}^{\prime -}_{W\bar{W}\gamma}\la {\bm 1}\,{\bm 2}\ra  \la 3\,{\bm 1} \ra \la 3\,{\bm 2}\ra  \la 3| p_{\bm 2}|4 \rb \lb 4\,3 \rb \right\}\,, \nn
\end{align}
which factorizes correctly in all channels. For the non-minimal coupling in the vertex $\hat{\mathcal{M}}({\bm 1}_W 3^{-}_{\gamma} {\bm q}_{\bar{W}})$ we obtain the same result with 
$\bd{...} \leftrightarrow \bb{...}$ and $3 \leftrightarrow 4$. Before discussing the non-factorizable contributions, let us analyse the HE of Eq.~(\ref{e:comptonbsm1}). Naively, we can just `unbold' the amplitude and  by the LG we can see that the term $\mathcal{C}_{W\bar{W}\gamma}^- $ contributes to amplitudes with longitudinal $W$'s, and  $\mathcal{C}_{W\bar{W}\gamma}^{\prime -} $ contributes at leading order to an amplitude with two transverse $W$ with the same polarization, which has the same structure as the gluon scattering with a $G^3$ insertion \cite{Dixon:1993xd}. 
Notice that the factorizable terms of the amplitude can also be obtained without BCFW, and just starting with an ansatz and requiring consistent factorization \cite{Chung:2018kqs}. However, both methods fail to  get the polynomial terms correctly. Since the term of Eq.~(\ref{e:cont}) is forbidden, the next candidate for contact interaction is 
\begin{align}
\frac{\mathcal{C}_{W\bar{W}\gamma \gamma}^-}{v^2\,\Lambda^2}\Big( \la 3\, {\bm 1}\ra\lb 4 \, {\bm 2}\rb + \la 3\, {\bm 2}\ra\lb 4 \, {\bm 1}\rb \Big)^2\,.
\end{align}
In this case, we know that the coefficient  $\mathcal{C}_{W\bar{W}\gamma \gamma}^-$ is  related to the 3-point vertex, as the information about the gauge symmetries of the theory are already in the map of Table~\ref{t:dim6}.

\subsection{$W\bar{W}hh$}
In the case of $\mathcal{M}({\bm 1}_W {\bm 2}_{\bar{W}}{\bm 3}_h {\bm 4}_h)$, one may expect a bad large-$z$ behaviour doing a BCFW shift due to the pure scalar amplitude with longitudinal W's and the Higgs. However, we may reconstruct the SM amplitude from the residues at the poles and requiring a well-behaved UV. The full amplitude can be written as 
\begin{align}
\label{e:amph}
\mathcal{M}({\bm 1}_W {\bm 2}_{\bar{W}}{\bm 3}_h {\bm 4}_h) = \frac{R_t}{t-m_W^2}+\frac{R_u}{u-m_W^2}+\frac{R_s}{s-m_h^2} + \mathcal{G}_{W\bar{W}hh} \la {\bm 1} {\bm 2}\ra \lb {\bm 1} {\bm 2} \rb\,,
\end{align}
where the form of the non-factorizable contribution is fixed by the $SU(2)$ LG and we will discuss how to obtain $\mathcal{G}_{W\bar{W}hh}$ in the following. First, we can easily calculate the residues by just `gluing' the corresponding 3-point amplitudes:
\begin{align}
R_t &=  \mathcal{M}({\bm 1}_W {\bm q}_{\bar{W}}{\bm 3}_h)\mathcal{M}(-{\bm q}_W {\bm 2}_{\bar{W}}{\bm 4}_h)|_{t\rightarrow m_W^2}=  \frac{(\mathcal{G}_{W\bar{W}h})^2}{v^2}\Big(\bd{\1{\bm q}}\bb{\1 {\bm q}}\Big)\Big(\bd{{\bm q}\2}\bb{{\bm q}\2}\Big)\,, \nn \\
R_s &=  \mathcal{M}({\bm 1}_W {\bm 2}_{\bar{W}}{\bm q}_h)\mathcal{M}(-{\bm q}_h {\bm 3}_{h}{\bm 4}_h)|_{s\rightarrow m_W^2}=(\mathcal{G}_{W\bar{W}h})(\mathcal{G}_{hhh}) \Big(\bd{\1{\bm 2}}\bb{\1 {\bm 2}}\Big) \,,
\end{align}
where ${\bm q}$ is the momentum in the corresponding factorization channel. The residue in the $u$-channel can then be obtained as $R_u = R_t\,({\bm 3}\leftrightarrow {\bm 4} )$. Notice that we have to sum over the $SU(2)$ indices of the internal particle such that
\begin{align}
\Big(\bd{\1{\bm q}}\bb{\1 {\bm q}}\Big)\Big(\bd{{\bm q}\2}\bb{{\bm q}\2}\Big) &=  \Big(\bd{\1^{J_1}{\bm q}^J}\bb{\1^{K_1} {\bm q}^{K}} + \bd{\1^{J_1}{\bm q}^K}\bb{\1^{K_1} {\bm q}^{J}} \Big)\Big(\bd{{\bm q}_J\2^{J_2}}\bb{{\bm q}_K\2^{K_2}}\Big) \nn \\
&= -m_1^2 \, \bd{\1^{J_1}{\bm 2}^{J_2}}\bb{\1^{J_1}{\bm 2}^{J_2}} + \la \1^{J_1}| \,p_{\bm q}\, |\2^{K_2} \rb \la \2^{J_2}| \,p_{\bm q}\, |\1^{K_1} \rb \,,
\end{align}
where $p_{\bm q}=p_{\1}+p_{\3}$ for the $t$-channel. Summing all the residues, we can use  $\mathcal{G}_i$ from Table~\ref{t:sm} and write Eq.~(\ref{e:amph}) as 
\begin{align}
\label{e:amph2}
\mathcal{M}({\bm 1}_W {\bm 2}_{\bar{W}}{\bm 3}_h {\bm 4}_h) =& \frac{(\mathcal{G}_{W\bar{W}h})^2}{v^2}\bigg[\frac{\la {\bm 1} | p_{\bm 3}|{\bm 1} \rb \la {\bm 2} | p_{\bm 4}|{\bm 2} \rb}{t-m_W^2}+\frac{\la {\bm 1} | p_{\bm 4}|{\bm 1} \rb \la {\bm 2} | p_{\bm 3}|{\bm 2} \rb}{u-m_W^2}\bigg] \\
\bigg[&-\frac{ (\mathcal{G}_{W\bar{W}h})^2m_W^2(2m_h^2 -s)}{v^2(t-m_W^2)(u-m_W^2)}-\frac{(\mathcal{G}_{W\bar{W}h})(\mathcal{G}_{hhh})}{ (s-m_h^2)} +\mathcal{G}_{W\bar{W}hh}\bigg]\la {\bm 1} {\bm 2}\ra \lb {\bm 1} {\bm 2} \rb\,. \nn
\end{align}
In the HE limit, we have that $\mathcal{M}_{\rm HE} \rightarrow  (\mathcal{G}_{W\bar{W}h})^2 s /(2v^2) + \mathcal{G}_{W\bar{W}hh} s$ 
and by unitarity we must have
\begin{align}
\mathcal{G}_{W\bar{W}hh} =-\frac{1}{2}\frac{(\mathcal{G}_{W\bar{W}h})^2}{v^2}= -\frac{2}{v^2}\,,
\end{align}
which reproduces the Feynman rules calculation. This should not be a surprise, since the 3- and 4-point contact interactions are connected by a gauge symmetry in the Lagrangian language, which means that the higher point contact interactions of this type are fixed by the coefficient of the 3-point amplitude. 

Moving to the SMEFT amplitude, the SM-like part gives  the same result as Eq.~(\ref{e:amph2}) with the replacement $\mathcal{G}_i \rightarrow\mathcal{C}_i$. However, we cannot use the HE constraint to obtain the new coefficient of the 4-point contact interaction $\mathcal{C}_{W\bar{W}hh}$, but it should be determined once we fixed the cubic ones. This channel can be particularly interesting to study the origin of the electroweak symmetry breaking. For example, one may explore the difference between SMEFT and HEFT, where the symmetry is linearly and non-linearly realized, respectively \cite{Falkowski:2019tft,Chang:2019vez,Alonso:2015fsp}. 

\section{Conclusion}
In this paper, we started the exploration of the massive SMEFT in terms of on-shell amplitudes using the formulation of massive spinors of  \cite{Arkani-Hamed:2017jhn}. Although the kinematic structure is fixed by Lorentz invariance, LG and Bose symmetry, it is not clear how the information about the UV symmetries and the relation between amplitudes should appear in the on-shell language. 

As a first step, we draw a map between the parameters of the SMEFT Lagrangian and the coefficients of the on-shell amplitudes using the Feynman rules derived in \cite{Dedes:2017zog}. We also discussed how the BCFW recursion relation can be applied for the massive SM $WW\gamma\gamma$ scattering. Similar calculations could also be done for other SM processes as e.g. $WWh\gamma$ or with a $Z$ boson instead. It also should be worthwile to develop a more systematic way to analyse the large-$z$ behaviour and in addition,  study general recursion relations.  For instance, a lot of progress was made using the soft and collinear limits in the massless and/or supersymmetric cases and one may ask if a similar approach can be useful also in the massive case, especially for the SMEFT. 

Another question is how the difference between a linear and a non-linear realization of the electroweak symmetry arises in this language. In other words, it would be interesting to see how HEFT and SMEFT are described at the on-shell level and if this can shed some light on the description of the spontaneous symmetry breaking \cite{rafacamila}. Moreover, we are just working at tree-level, but there are several avenues to pursue loop calculation also with massive particles. We hope that this may be an initial step towards a deeper understanding of the intersection of on-shell amplitudes and SMEFT.

%------------------------------------------%
\subsection*{Acknowledgments}
We thank Adam Falkowski for the scientific guidance and encouragement on this project. We also thank Andreas Helset, Brian Henning, Sophie Renner, Dave Sutherland and William Shepherd for fruitful discussions and for comments that greatly improved the manuscript. The work of CSM and RA was supported
by the Alexander von Humboldt Foundation, in the framework of the Sofja Kovalevskaja Award 2016, endowed by the German Federal Ministry of Education and Research and also supported by  the  Cluster  of  Excellence  ``Precision  Physics,  Fundamental
Interactions, and Structure of Matter" (PRISMA$^+$ EXC 2118/1) funded by the German Research Foundation (DFG) within the German Excellence Strategy (Project ID 39083149).
%------------------------------------------%

%------------------------------
\appendix

%------------------------------

\section{Review of massless and massive spinors}
\label{sec:review}
In the massless case, the Lorentz algebra $SL(2,\mathbb{C})$ can be decomposed into $SU(2)\times SU(2)$ and the momentum bi-spinor is written as
\begin{align}
p_{\alpha \dot{\beta}}&=\lambda_{\alpha}\tilde{\lambda}_{\dot{\beta}}\equiv |p \rangle_\alpha \lbrack p |_{\dot{\beta}}\,, \nonumber  
%p^{\dot{\alpha} \beta}&=\tilde{\lambda}^{\dot{\alpha}}\lambda^{\beta} \equiv |p \rbrack^{\dot{\alpha}} \langle p |^{\beta}\,,
\end{align}
which is the contraction of 4-momentum with $\sigma^{\mu}_{\alpha \dot{\beta}}=(1,\vec{\sigma})$ and $\bar{\sigma}_{\mu}^{\dot{\alpha}\beta}=(1,-\vec{\sigma})$ where $\sigma^i$ are the Pauli matrices. The polarization vectors for a massless spin-1 particle is given by
\beq
\label{e:masslesspol}
	\eps^\m_+ = \frac{ \bdp{\zeta}{\sig^\m}{\lam} }{ \sqrt{2}\bd{\lam\zeta} }\,, \qquad \eps^\m_- = \frac{ \bdp{\lam}{\sig_\m}{\zeta} }{ \sqrt{2}\bb{\lam\zeta}\, }\,, 
\eeq
where $\lam$ represents the particle spinor and $\zeta \neq \lambda$ an arbitrary reference spinor. The little group, i.e. the group of transformations that leave the on-shell momenta invariant, is $U(1)$ for massless particles which means that $|p\ra \rightarrow t\,|p\ra$ and $|p\rb \rightarrow t^{-1}\,|p\rb$.

For the massive case $\det \p_{\alpha \dot{\beta}} =m^2$ and as the little group is $SU(2)$ we can decompose the momentum in terms of 4 spinors as~\cite{Arkani-Hamed:2017jhn}
\begin{align}
\p_{\alpha \dot{\beta}} &=  \bchi_{\alpha}^I \epsilon_{IJ} \tilde{\bchi}_{\dot{\beta}}^J \equiv \epsilon_{IJ}| \p \rangle^I_{\alpha}\lbrack \p |_{\dot{\beta}}^J, \nonumber
%\p^{\dot{\alpha} \beta} &=  -\bchi^{\dot{\alpha}\,I} \epsilon_{IJ} \tilde{\bchi}^{\beta \, J} \equiv -\epsilon_{IJ}| \p \rbrack^{I\,\dot{\alpha}}\langle \p|^{\beta \, J}\,,
\end{align}
where $\alpha, \beta$ are $SL(2,\mathbb{C})$ indices, $I,J=\{1,2\}$ are $SU(2)$ indices. For clarity, we use {\bf bold} to indicate the quantities related to massive particles although it will be clear from the $SU(2)$ indices. For a massive spin-1 particle, we can define a polarization tensor as 
\begin{align}
\label{e:massivepol}
	[\eps^\m]^{JK}& =\frac{1}{\sqrt{2}m} \bdp{\bchi^J}{\sig^\m}{\bchi^K}\, ,\nonumber 
%	\\ [\bar{\eps}^\m]^{JK}& =-\frac{1}{\sqrt{2}m} \bdp{\bchi_J}{\sig^\m}{\bchi_K}
\end{align}
where the symmetrization on $SU(2)$ indices is implicit.

\subsection{High energy limit}
For $\p^{\mu}=(E, Ps_{\theta}c_{\phi},Ps_{\theta}s_{\phi}, Pc_{\theta})$ we have $E^2-P^2=m^2$ and the explicit solution for the spinors is given by,
\begin{align}
|\p \rangle_{\alpha}^I = \begin{pmatrix}
\sqrt{E-P}\,c & -\sqrt{E+P}\,s^* \\
\sqrt{E-P}\,s & \sqrt{E+P}\,c
\end{pmatrix} \,,
\\
\lbrack\p |_{\dot{\beta}}^J = \begin{pmatrix}
\sqrt{E-P}\,c & -\sqrt{E+P}\,s \\
\sqrt{E-P}\,s^* & \sqrt{E+P}\,c
\end{pmatrix} \,, \nonumber
\end{align}
which can be written as
\begin{align*}
\label{e:massspinors2}
&|\p \rangle_{\alpha}^I = \sqrt{E-P}\zeta_{\alpha}^{+}(\theta)\otimes\zeta^{-\,I}\!-\!\sqrt{E+P}\zeta_{\alpha}^{-}(\theta)\otimes\zeta^{+\,I},\nonumber\\
&\lbrack\p |_{\dot{\beta}}^J= \sqrt{E+P}\tilde{\zeta}_{\dot{\beta}}^{-}(\theta)\otimes\zeta^{+\,J}\!+\!\sqrt{E-P}\tilde{\zeta}_{\dot{\beta}}^{+}(\theta)\otimes\zeta^{-\,J},
\end{align*}
making clear that the two indices belong to two different spaces. The high energy limit can be obtain taking $m/E \ll 1$. It is useful to define
\bea
	\lam &=& \sqrt{E+P} \,\zeta^-(\theta), \quad \lamt = \sqrt{E+P}\,\tilde{\zeta}^+(\theta), \\ \nonumber
	\eta &=& \sqrt{E-P} \,\zeta^+(\theta), \quad \etat = \sqrt{E-P}\,\tilde{\zeta}^-(\theta),
\eea
from which follows immediately that $\eta$ and $\tilde{\eta}$ are suppressed by a mass term while $\lam$ and $\lamt$ are the massless spinors. 
With these, we can write the massive spinors as:
\bea
				\bchi^J_\al  &=& - \lam_\al \zeta^J_+ + \eta_\al \zeta^J_-, \\ \nonumber
				\tilde{\bchi}_{\btp}^J &=& \lamt_{\btp} \zeta^J_- + \tilde{\eta}_{\btp}\tilde{\zeta}^J_+\,. 
\eea
We can see that in the high energy limit, these spinors collapse to the $\lam$'s spinors since $\eta$ and $\tilde{\eta}$ are mass suppressed.
\beq
	\lim_{m\to 0} \bchi^J_\al = - \lam_\al \zeta^J_+  \quad \text{and} \quad \lim_{m\to 0} \tilde{\bchi}_{\btp}^J = \lamt_{\btp} \zeta^J_- \,.
\eeq

\subsection{3-point massive amplitudes}
Any amplitude with a mix of massive and massless legs can be written by stripping the massive spinors and constructing the 
rest with massless little group restriction. Here we are following closely Ref. \cite{Arkani-Hamed:2017jhn}.

\subsubsection{Two massless + One massive  particles }
An amplitude with a massive particle with spin $s$ and mass $m$ and two massless particles with helicity $h_i$ can be decomposed stripping out the $SU(2)$ massive
indices as
\begin{align}
\mathcal{M}(\bm{1}^{I_{1} \cdots \,I_{2s}}\, 2^{h_2}\, 3^{h_3}) &= M^{\{\alpha_1 \cdots \alpha_{2s}\}}\,\bm{\chi}_{\alpha_1}^{I_{1}} \cdots \,\bm{\chi}_{\alpha_{2s}}^{I_{2s}}= \tilde{M}^{\{\dot{\alpha}_1 \cdots \dot{\alpha}_{2s}\}}\,\bm{\tilde{\chi}}_{\dot{\alpha}_1}^{I_{1}} \cdots \,\bm{\tilde{\chi}}_{\dot{\alpha}_{2s}}^{I_{2s}}\,,
\end{align}
with the stripped amplitude written as
\begin{align}
&M^{\{\alpha_1 \cdots \alpha_{2s}\}} =g\, \lb 2\,3 \rb^{s+h_2+h_3}\left(\lambda_2^{s+h_3-h_2}\lambda_3^{s-h_3+h_2}\right)^{\{\alpha_1 \cdots \alpha_{2s}\}}\,,
\end{align}
or
\begin{align}
&\tilde{M}^{\{\alpha_1 \cdots \alpha_{2s}\}} = g\, \la 2\,3 \ra^{s+h_2+h_3}\left(\tilde{\lambda}_2^{s+h_3-h_2}\tilde{\lambda}_3^{s-h_3+h_2}\right)^{\{\alpha_1 \cdots \alpha_{2s}\}}\,,
\end{align}
where $I_a=1,2$ with $a=1\cdots 2s$ are $SU(2)$ indices and dimensional analysis gives $\lb g \rb=1-(3s +h_2+h_3)$. A non-trivial constrain on the amplitude comes from that the expression above only exist if $|h_3-h_2|\leq s$.
\subsubsection{One massless + Two massive particles }
An amplitude with two massive particles with spin $s_1,s_2$ and mass $m_1,m_2$ coupling with one massless particle with helicity $h_3$ can be decomposed as
\begin{align}
&\mathcal{M}(\bm{1}^{I_{1} \cdots \,I_{2s_1}}\, \bm{2}^{J_{1} \cdots \,J_{2s_2}}\, 3^{h_3}) = M^{\{\alpha_1 \cdots \alpha_{2s_1}\}\{\beta_1 \cdots \beta_{2s_2}\}}\,\bm{\chi}_{\alpha_1}^{I_{1}} \cdots \,\bm{\chi}_{\alpha_{2s_1}}^{I_{2s_1}}\, \,\bm{\psi}_{\beta_1}^{J_{1}} \cdots \,\bm{\psi}_{\beta_{2s_1}}^{J_{2s_2}}\,.
\end{align}
In the case that $m_1 \neq m_2$, the spinors $u^{\alpha} \equiv \lambda_3^{\alpha}$ and $v_{\alpha} \equiv p_{\bm{1}}(\sigma_{\alpha \dot{\alpha}}\tilde{\lambda}_3^{\dot{\alpha}})/\sqrt{m_1\,m_2}$ span the entire $2D$ spinor space allowing the amplitude to be written as
\begin{align}
&M^{\{\alpha_1 \cdots \alpha_{2s_1}\}\{\beta_1 \cdots \beta_{2s_2}\}} =\sum_{i=1}^{C} g_i \left(u^{s_1+s_2+h_3} v^{s_1+s_2-h_3} \right)^{\{\alpha_1 \cdots \alpha_{2s_1}\}\{\beta_1 \cdots \beta_{2s_2}\}}\,, \nonumber
\end{align}
where $C=2\text{min}(s_1,s_2)+1$ is the number of different partitions of the two groups of $SL(2,\mathbb{C})$ indices.

In the case that $m_1 = m_2$ we need to define the auxiliary objects,
\begin{align}
x\equiv \frac{\langle \zeta |p_{\bm{1}} |3 \rbrack}{\langle 3\, \zeta  \rangle}, \qquad x^{-1} \equiv \frac{\langle 3 |p_{\bm{1}} |\tilde{\zeta} \rbrack}{\lbrack \tilde{\zeta}\, 3 \rbrack},
\end{align}
where $3$ is the label of the massless particle. The amplitude is then given by
\begin{align}
&M^{\{\alpha_1 \cdots \alpha_{2s_1}\}\{\beta_1 \cdots \beta_{2s_2}\}} = \sum_{i=|s_1-s_2|}^{s_1+s_2} g_i\, (m\,x)^{h+i}\left( \lambda_3^{2i} \varepsilon ^{s_1+s_2-i} \right)^{\{\alpha_1 \cdots \alpha_{2s_1}\}\{\beta_1 \cdots \beta_{2s_2}\}}\,. 
\end{align}
where $[g_i]=1-i-s_1-s_2$.

\subsubsection{Three massive particles}
For amplitudes with three massive particles, one should write all possible terms consistent with the little group. These are not so many structure as $\la \bm{i}\,\bm{i}\ra =\lb \bm{i}\,\bm{i}\rb =0$ after the symmetrization of the $SU(2)$ indices. Also, structures with a momentum insertion can be rewritten in terms of contraction of angle and square brackets with the use of Schouten identities.

\subsection{Useful identities}
In the following we are going to list useful identities for the manipulation of the massive amplitudes. For $\sigma_\m$ and $\bar{\sigma}_\m$ we can write that
\beq
	\sig^\m_{\al\alp} = \eps_{\al\bt}\eps_{\alp\btp}\sigb_\m^{\bt\btp}\,, \qquad 
	\sigb_\m^{\al\alp} = \eps^{\al\bt}\eps^{\alp\btp}\sig^\m_{\bt\btp}\,.
\eeq
Therefore, the Fierz identities for $\sigma$ matrices can be written as
\begin{flalign}
&	[\sig^\m]_{\al\alp} [\sig^\n]_{\al\alp} = 2 \eps_{\al\bt}\eps_{\alp\btp}\,, \nonumber \\
&	[\sig^\m]_{\al\alp} [\sig^\n]_{\al\alp} 
	= \frac{1}{2}( [\sig^\m]_{\al\btp} [\sig^\n]_{\btp\alp} + [\sig^\n]_{\al\btp} [\sig^\m]_{\btp\alp}  2\eta^{\m\n}\eps_{\al\bt}\eps_{\alp\btp} + i \eps^{\m\n\rho\sig} [\sig^\rho]_{\al\btp} [\sig^\sigma]_{\btp\alp} ) \,. \nonumber
\end{flalign}
With these identities, one can write the Fierz identities for the spinors (omitting the $SU(2)$ indices):
\begin{flalign}
&	\bdp{\1}{\sig_\m}{\2}\bdp{\3}{\sig^\m}{\4} = -2 \bd{\1\3}\bb{\2\4}\,, \nonumber \\
	&\bdp{\1}{q\sig}{\2}\bdp{\3}{q\sig}{\4}     =	\bdp{\1}{q\sig}{\4}\bdp{\3}{q\sig}{\2} +\, q^2 \bd{\1\3}\bb{\2\4}\,,  \nonumber\\
&	\bdp{\1}{\sig^\m}{\2}\bdp{\3}{\sig^\n}{\4} =  
	\frac{1}{2}\Big\{\! \bdp{\1}{\sig^\m}{\4}\bdp{\3}{\sig^\n}{\2} \!+\! \bdp{\1}{\sig^\n}{\4}\bdp{\3}{\sig^\m}{\2}  \nonumber\\
&\qquad \qquad	\qquad \quad ~ + 2\eta^{\m\n} \bd{\1\3}\bb{\2\4} + i \eps^{\m\n\rho\sig} \bdp{\1}{\sig^\rho}{\4}\bdp{\3}{\sig^\sig}{\2} \Big\}  \,.\nonumber
\end{flalign}
The Schouten identity is given by
\begin{flalign}
\bd{\1\2}\bd{\3\4} - \bd{\1\3}\bd{\2\4} + \bd{\1\4}\bd{\2\3}=  0\,,
\end{flalign}
and similarly for the square brackets. We can also write this directly in terms of polarization vector contractions:
\begin{flalign}
	&\bdp{\1}{\sig^\m}{\2
	}\bdp{\2}{\sig^\n}{\1} =  (m_1m_2)\{\eps^\m_1 \eps^\n_2 + \eps^\m_1 \eps^\n_2 + \eta^{\m\n} (\eps_1 \cdot \eps_2) + i \eps^{\m\n\rho\sig}\eps_{1\rho}\eps_{2\sig}\}\,, 
\end{flalign}
and, for three massive vectors, we can write 
\begin{flalign}
\,&(\eps_i \cdot \eps_j)(\eps_k \cdot p) = -\left(\frac{\bd{\ib\jb}\bb{\ib\jb}\bdp{\kb}{p}{\kb}}{\sqrt{2}m_im_jm_k}\right)\,,\nn \\
\,&(\eps_k \cdot p)(\eps_i \cdot p_j)(\eps_j\cdot p_i) =- \frac{\bdp{\kb}{p}{\kb}}{2\sqrt{2}m_k}\left(\bd{\ib\jb}\bd{\ib\jb} + \bb{\ib\jb}\bb{\ib\jb} - \frac{M_k \bd{\ib\jb}\bb{\ib\jb}}{m_im_jm_k}\right)\,,\nn \\ 
	\,&i(\eps_k\cdot p)\varepsilon^{\mu\nu\rho\sigma} p_j^\mu p_i^\nu \eps_j^\sigma \eps_i^\rho = -\frac{\bdp{\kb}{p}{\kb}}{2\sqrt{2}m_k}\Big(\bd{\ib\jb}\bd{\ib\jb} - \bb{\ib\jb}\bb{\ib\jb}\Big)\,,
\end{flalign}
where $M_k = m^2_k - m^2_i - m^2_j$.
 %%%%%%%%%%%%%%%%%%%%%%%%%%%%%%%%%%%%%%%%%%%%%%%%%%%

 %\bibliographystyle{apsrev4-1.bst}
%%%%%%%  \bibliographystyle{utphys.bst}
\bibliographystyle{JHEP}
  \bibliography{refs}

\end{document}